\newcommand{\beq}{\begin{equation}}
\newcommand{\eeq}{\end{equation}}
\newcommand{\eq}[1]{\begin{align}#1\end{align}}
\newcommand{\lf}{\left}
\newcommand{\rf}{\right}
\newcommand{\nt}{\notag}

\documentclass[10pt, prb,aps,twocolumn,superscriptaddress]{revtex4-1}
\usepackage{graphicx} 
\usepackage{bm}
\usepackage{booktabs}
\usepackage{amsmath,amsthm,amssymb}
\usepackage{color}

\begin{document}

\title{
Photovoltaic anomalous Hall effect in line-node semimetals}

\author{Katsuhisa Taguchi}
\affiliation{Department of Applied Physics, Nagoya University, Nagoya, 464-8603, Japan}

\author{Dong-Hui Xu}
\affiliation{Department of Physics, Hong Kong University of Science and Technology, Clear Water Bay, Hong Kong, China}

\author{Ai Yamakage}
\affiliation{Department of Applied Physics, Nagoya University, Nagoya, 464-8603, Japan}
\affiliation{Institute for Advanced Research, Nagoya University, Nagoya, 464-8601, Japan}

\author{K. T. Law}
\affiliation{Department of Physics, Hong Kong University of Science and Technology, Clear Water Bay, Hong Kong, China}

\date{\today }
%

\begin{abstract}

We theoretically study the circularly polarized light-induced Floquet state in line-node semimetals with time-reversal symmetry and inversion symmetry.
It is found that the Floquet state can show the photovoltaic anomalous Hall effect when an applied circularly polarized light gaps out the line node in the bulk and leave Weyl point nodes.
The Hall conductivity is sensitive to the location of Fermi level: When the Fermi level locates at the node, the Hall conductivity depends on the radius of line node and is nearly independent of the intensity of light. 
Away from the line node, the Hall conductivity is dependent on the intensity of light. Such a sensitive Fermi-level dependence of the Hall conductivity in the presence of a weak laser intensity can have applications in phototransistors based on thin films of line-node semimetals.
\end{abstract}

\maketitle

\section{Introduction}
Topological matters have attracted enormous attention in recent years because they can host exotic edge or surface states protected by nontrivial bulk topology.
Topological insulators, as well known examples, are characterized by a fully insulating gap in the bulk and symmetry protected metallic states on the boundaries \cite{rf:rmp1,rf:rmp2}.
Nowadays, the research interest in topological materials has moved in part from insulators to semimetals as semimetals with topologically non-trivial Fermi surfaces can also support robust surface states. In general, the bulk band structures of topological semimetals possess point or line nodes in momentum space \cite{rf:Balents,rf:Chiu2}.
3D Dirac semimetal is one class of topological semimetals with four-fold degenerate point nodes and the electrons have linear dispersions near the Dirac nodes.
Na$_3$Bi \cite{rf:zkliu1} and Cd$_3$As$_2$ \cite{rf:zkliu2,rf:neupane} have been experimentally confirmed to be topological Dirac semimetals, where the Dirac nodes are protected by crystalline symmetry. If inversion or time-reversal symmetry is broken, each Dirac node splits into two Weyl nodes, separated in the momentum space, and the systems become Weyl semimetals \cite{rf:xgwan}. Weyl nodes with distinct chiralities lead to a variety of exotic measurable consequence, such as Fermi arc surface states and chiral anomaly.
Weyl semimetals have become a hot topic because the real Weyl semimetal materials has been theoretically proposed \cite{rf:smhuang,rf:xdai} and experimentally discovered in the inversion symmetry breaking TaAs class crystal \cite{rf:hasan,rf:hding,rf:hasan2,rf:lxyang}.

Differently from the Dirac/Weyl semimetals where the conduction band touches the valence band at discrete points in the momentum space, the conduction and valence bands in topological line-node semimetals touch each other on closed lines and symmetry protected drumhead surface states emerge \cite{rf:Balents, rf:Chiu, rf:Fang, rf:Chan2016, rf:Chiu2}.
Recently, several theoretical proposals \cite{rf:Mullen, rf:cava1,rf:cava2,rf:rappe,rf:xhu,rf:Bian,rf:hasantnl2,rf:Schoop, rf:Weng, rf:Huang} and
experimental studies \cite{rf:Bian,rf:Schoop,rf:Xie,rf:Neupane2016,rf:Wu,rf:Hu,rf:Tanaka16} appeared on line-node materials.
Since the dimension of line nodes is different from that of point nodes in Dirac/Weyl semimetals, one expects that line-node semimetals exhibit new topological transport and response phenomena characteristic to line nodes.
Several works, indeed, have shown novel phenomena in line-node semimetals, e.g., the minimal conductivity \cite{rf:Balents, rf:Mullen}, quantized Hall conductivity \cite{rf:Mullen}, flat Landau level \cite{rf:Rhim}, plasmons \cite{rf:Rhim2, rf:Yan}, and electric polarization and orbital magnetization \cite{rf:Ramamurthy}.
One way to realize nontrivial transport phenomena in semimetals is to irradiate the materials by light. Dirac/Weyl semimetals in the presence of circularly polarized light have been theoretically studied within the framework of Floquet theory \cite{rf:Oka15,rf:Chan15,rf:Taguchi16}. The circularly polarized light couples to electrons in a nontrivial form: the angular momentum of the incident light interacts with that of an electron which can break the time reversal symmetry. The anomalous Hall effect due to the light-induced interaction, which is called photovoltaic anomalous Hall effect \cite{rf:Oka09}, occurs when the Fermi level is located near the Weyl nodes. 

In this paper, we study the transport phenomena of line-node semimetals in the presence of circularly polarized light within the framework of Floquet theory. We find that the light-induced interaction in line-node semimetals, unlike that in the Dirac/Weyl semimetals \cite{rf:Oka15,rf:Chan15,rf:Taguchi16}, has strong dependence on the direction of incident light and highly anisotropic. We present that photovoltaic anomalous Hall effect in line-node semimetals with time reversal and inversion symmetries, such as in Ca$_3$P$_2$ \cite{rf:Xie}, can be generated by applying circularly polarized light.
We calculate the Hall conductivity $\sigma_{zx}^{\textrm{AHE}}(\epsilon_{\textrm{F}}) $  as functions of Fermi level $\epsilon_{\textrm{F}}$ and temperature. It is found that for $\epsilon_{\textrm{F}}=0$, which means the Fermi level is located at line node, the Hall conductivity depends on the radius of the line-node but it does not depend on the strength of the light-induced interaction. For  $\epsilon_{\textrm{F}}\neq0$, on the other hand, the Hall conductivity depends both the radius of the line-node and the strength of light-induced interaction.
As a result, the magnitude of $\sigma_{zx}^{\textrm{AHE}}(\epsilon_{\textrm{F}}) $ is sensitive to $\epsilon_{\textrm{F}}$ in the presence of a weak intensity of the incident light.


\section{Model} 
We consider a line-node semimetal with time-reversal and spatial-inversion symmetries which can be described by a minimal two-band model.
In the absence of applied electromagnetic field,
the low-energy effective model Hamiltonian of line node semimetal is expressed by \cite{rf:rappe,rf:Weng2015,rf:xhu}
\eq{
	\label{eq:2-1} 
H_0  & =  \sum_{\bm{k}}  \psi^\dagger_{\bm{k}}
	\mathcal{H}_0 \psi_{\bm{k}},\\
\mathcal{H}_0 (\bm{k})& =
	\lf( \frac{\hbar^2 k^2 }{2m}- m_0 \rf) \tau^z s^0
	+ vk_z  \tau^y s^0 - \epsilon_{\textrm{F}} \tau^0 s^0,
}
where $m$ is the effective mass of electron,  $\psi_{\bm{k}} = (\psi_{\uparrow, +} \  \psi_{\uparrow, -} \ \psi_{\downarrow, +} \  \psi_{\downarrow, -} )^T$
is the annihilation operator with spin up ($\uparrow$), down ($\downarrow$), and even- ($+$), odd- ($-$) parity orbitals under the mirror reflection with respect to the $z=0$ plane,
and $k^2 = k_x^2 + k_y^2+k^2_z$ is considered.
$s^{x,y,z}$ ($\tau^{x,y,z}$) and $s^0$ ($\tau^0$) are the Pauli matrices and the identity matrix in the spin (orbital) space, respectively.
A line node appears on the circle of $k_x^2 + k_y^2 = 2mm_0/\hbar^2$ and $k_z=0$ when the bands are inverted ($m_0 > 0$).
$m_0$ and $v$ are parameters corresponding to the radius of the line node and the velocity of the $z$ direction, respectively.
%

We further take into account the electromagnetic fields for line-node semimetals. 
The total Hamiltonian is obtained by Peierls substitution, $\bm{k} \rightarrow \bm{k}-e\bm{A}/\hbar$. 
Keeping the first order term expansion about vector potential $\mathbf{A}$, the total Hamiltonian is approximately given by
\eq{\label{eq:2-2} 
H & = H_{0} + H_{\textrm{em}},
	\\ \label{eq:2-3} 
H_{\textrm{em}} & = - \sum_{\bm{k}}  \psi^\dagger_{\bm{k}} \bm{j} \cdot \bm{A} \psi_{\bm{k}}.
}
where the charge current $\bm{j}$ is represented by
\eq{
\label{eq:2-3-a}
	\bm{j} 
	    & = \frac{e\hbar}{m} \lf(\bm{k} - \frac{e}{2\hbar} \bm{A}\rf)  \tau^z s^0
	 - \frac{ev}{\hbar} \tau^y s^0  \bm{e}_z,
}
with $e<0$ being the elementary charge of an electron and $\bm{e}_z$ is the unit vector along the $z$-direction.
The electric field $\bm E$ is obtained from the spatially-uniform vector potential $\bm{A}$; $\bm{E} = -\partial_t \bm{A}$.

It is noted that in Sec. \ref{sec:Floquet},
$\bm{A} = \bm A_{\rm L}(t)$ and $\bm{E} = \bm E_{\rm L}$ denote the vector potential and electric field of the incident light, respectively.
In Sec. \ref{sec:PAHE}, to consider the anomalous Hall effect,
a DC electric field ($\bm{E}_{\textrm{dc}}$) is also included:
$\bm{E} \equiv \bm{E}_{\rm L} + \bm{E}_{\textrm{dc}}$ 
and $\bm{A} \equiv \bm{A}_{\rm L} + \bm{A}_{\textrm{dc}}$.

\section{Floquet states}
\label{sec:Floquet} 
Based on the model Hamiltonian of Eqs. (\ref{eq:2-1})-(\ref{eq:2-3}), we consider the Floquet state in line-node semimetals in the standard manner \cite{rf:Oka15,rf:Chan15,rf:Taguchi16,rf:Oka09}.
Consider time-dependent vector potential $\mathbf{A}(t)$ is a periodic function with period of $2\pi/\Omega$, then the total Hamiltonian eqn. (\ref{eq:2-2}) is also periodic as
$\mathcal{H}(t) = \mathcal{H}(t+ 2\pi/\Omega)$ with the frequency of the incident light  $\Omega$.
Below, $\bm{A}_{\rm L}$ is assumed to be the monochromatic frequency of the coherent light.
Then, the wave function of the Schr\"odinger equation $i\partial_t \psi (t) = H \psi (t)$ is given by
$\psi(t)  = \sum_{u} \phi_u e^{-i( \epsilon/\hbar +u\hbar\Omega)t }$,
where $\epsilon$ is the Floquet quasi-energy and $u$ takes all integers.
From the Schr\"odinger equation, the Floquet equation,
$\sum_n \mathcal{H}_{u,n} \phi_n = ( \epsilon + u \hbar \Omega) \phi_u$ is obtained.
Here,  $\mathcal{H}_{u,n}  \equiv \frac{1}{\Omega/(2\pi)} \int_0^{\Omega/(2\pi)} \mathcal{H}(t) e^{i(u-n)\Omega t} dt + u \hbar \Omega \delta_{un}$ is the block Hamiltonian in the Floquet state.
From Eq. (\ref{eq:2-1}), the diagonal term becomes
$\mathcal{H}_{u,u} =  u \hbar \Omega + \frac{1}{\Omega/(2\pi)} \int_0^{\Omega/(2\pi)} \mathcal{H}(t) dt =  u \hbar \Omega + \mathcal{H}_0 + \frac{e^2}{2m} |A_{\rm L}|^2 \tau^z s^0 $ and the off-diagonal terms are
$\mathcal{H}_{u,u+1} = \mathcal{H}_{u+1,u}^\dagger = - \frac{1}{\Omega/(2\pi)} \int_0^{\Omega/(2\pi)} \bm{j}\cdot\bm{A}_{\rm L}(t) e^{-i\Omega t}dt$.
The other off-diagonal terms are identically zero.
Each solution of the Floquet equation is regarded as a periodic steady state.

Now we focus on the effective Hamiltonian $\tilde {\mathcal H}_{0,0}$, integrating out the higher-energy states $\phi_{n \geq 1}$.
$H_{\rm em}$ renormalizes the parameters and introduces some new terms into $\tilde{\mathcal H}_{0,0}$.
From the symmetry considerations, one finds that $k_y \tau^x s^0$ (or $k_x \tau^x s^0$) is the only time-reversal-symmetry-breaking term induced by $H_{\rm em}$ in $\tilde {\mathcal H}_{0,0}$ without spin-orbit interaction (see Appendix \ref{symmetry} for details).
This term gives rise to the photovoltaic anomalous Hall effect, as discussed in Sec. \ref{sec:PAHE}.
In the following, we actually derive the light-induced terms in the effective Hamiltonian within the second-order perturbation theory.

It is assumed that for a perturbation theory, the energy scale of the incident light ($\hbar\Omega$) is larger than the width of the energy scale in Eq. (\ref{eq:2-1}).
Then, the off-diagonal term is regarded as a perturbation for $\mathcal{H}_0$, and
the effective Hamiltonian $\mathcal{H}_{\textrm{eff}}=\tilde {\mathcal H}_{0,0}$ in the periodic steady state is given by
\eq{
 \label{eq:Floquet-5} 
\mathcal{H}_{\textrm{eff}} & = \mathcal{H}_0+ \frac{e^2 |A_{\rm L}|^2}{2m} \tau^z s^0 + \frac{[\mathcal{H}_{0,-1}, \mathcal{H}_{0,1}]}{\hbar\Omega}  + \mathcal{O}(A_{\rm L}^4).
}
The first and second terms come from the diagonal term $\mathcal H_{0,0}$ and the third term is the second-order correction owing to the off-diagonal terms.
It is noticed that the second term and third term are newly added in the Hamiltonian.
Below, we consider the detail of these terms and the physical meaning, which are summarized in Table \ref{table1}.

\begin{table}
\centering
\caption{Effects of circularly polarized light along the $z$, $x$, and $y$ axes in a semimetal hosting a line node on the $k_z=0$ plane. PAHE stands for the photovoltaic anomalous Hall effect.}
\begin{tabular}{ccc}
 \hline\hline
 Axis & Induced term & PAHE
 \\
 \hline
 $z$ & $\delta m_0 \tau^z s^0$ & 0
 \\
 $x$ & $\delta m_0 \tau^z s^0$, $-\mathcal L_x k_y \tau^x s^0$ & $\sigma_{yz}^{\rm AHE}$
 \\
 $y$ & $\delta m_0 \tau^z s^0$, $-\mathcal L_y k_x \tau^x s^0$ & $\sigma_{xz}^{\rm AHE}$
 \\
 \hline\hline
\end{tabular}
\label{table1}
\end{table}

\subsection{ Light propagation along the $z$ axis}\label{sec:Floquet-z} 
When the light is along the $z$ axis, $\bm{A}_{\textrm{L}}$ is given by
\eq{
\bm{A}_{\textrm{L}}  = A_{\rm L} (\cos \Omega t, \sigma_{\textrm{L}}^z \sin\Omega t, 0),
}
where
$ A_{\rm L} \equiv i E_{\rm L}/\Omega$ and $E_{\rm L}$ are the magnitudes of the vector potential and the electric field of the light, respectively.
The spin angular momentum of light $\sigma_{\textrm{L}}^z=\pm1$ indicates the chirality of right-handed and left-handed circularly polarized light.
Then, $[\mathcal{H}_{0,-1}, \mathcal{H}_{0,1}]$ vanishes and the effective Hamiltonian is given by
\begin{align}
\mathcal{H}_{\textrm{eff}}
	 & = \mathcal{H}_0  + \delta m_0 \tau^z s^0,
 \label{Heff-z}
\end{align}
with
\begin{align}
\label{eq:Floquet-10} 
\delta m_0 & =\frac{e^2 E_{\rm L}^2 }{2m\Omega^2}.
\end{align}
Thus, there is no time-reversal-symmetry-breaking term in $\mathcal{H}_{\textrm{eff}}$.
It is found that in Eq. (\ref{Heff-z}) the induced term is proportional to $\tau^z s^0$ and its sign is always positive $\delta m_0 > 0$,
hence this term plays a role to decrease $m_0$ of Eq. (\ref{eq:2-1}) by $\delta m_0$.
In other words, $m_0$ is renormalized into $\bar m_0$;
\eq{
 \label{eq:Floquet-12} 
m_0 \to \bar m_0 = m_0 - \delta m_0.
} It means that applying light along the $z$ direction just changes the radius of line node, assuming $\delta m_0$ is smaller than $m_0$

A line node may induce the quasi topological response of the electric polarization $P_z$ and the orbital magnetization $M_z$ along the $z$ direction, which are proportional to $m_0$ \cite{rf:Ramamurthy}.
This implies that $P_z$ and $M_z$ would be controlled by the circularly polarized light along the $z$ direction.
It is also found that
the change of $P_z$ and $M_z$ are independent of the chirality of the incident light.

\subsection{ Light along the $y$ axis}\label{sec:Floquet-y} 
When the incident light propagates along the $y$ axis,
a new term, which breaks time-reversal symmetry, is induced in addition to the $\delta m_0$ term, as shown below.
$\bm{A}_{\rm L}$ is represented by
\eq{
 \label{eq:Floquet-6} 
\bm A_{\rm L}
	& = A_{\rm L} ( \sigma_{\textrm{L}}^z \sin\Omega t,0, \cos\Omega t ),
}
The off-diagonal term $\mathcal{H}_{0,-1} (= \mathcal{H}_{0,1}^\dagger )$ is given by
\eq{
& \mathcal{H}_{0,-1}     = \frac{i eE_{\rm L}}{2\hbar \Omega}
		\begin{pmatrix}
 		&- \frac{\hbar^2}{m} [\sigma_{\textrm{L}}^z  i k_x    + k_z]&
		 &i v&
		\\
		&-i v&
		&\frac{\hbar^2}{m}  [\sigma_{\textrm{L}}^z i k_x    + k_z]&
		\end{pmatrix},
}
and the third term of Eq. (\ref{eq:Floquet-5}) becomes
\eq{
\frac{1}{\hbar \Omega}[\mathcal{H}_{0,-1}, \mathcal{H}_{0,1}]=
 -\frac{v e^2 E_{\rm L}^2 }{m\hbar \Omega^3} \sigma_{\textrm{L}}^z k_x \tau^x  s^0.
}
The above term is represented by the form $i \bm{\mathcal{E}} \times  \bm{\mathcal{E}}^* = |\bm{\mathcal{E}}|^2 \sigma_{\textrm{L}}^z \bm{q} $, which indicates the direction of the light propagation and the chirality of the incident polarized light, where $\bm{q}$ is the unit vector of the incident light and $\bm{\mathcal{E}}$
and $\bm{\mathcal{E}}^*$ are the complex vector of the electric field and its complex conjugate, respectively.
When the light is along the $y$ axis, the complex vector is given by
$\bm{\mathcal{E}} = E_{\rm L} (i, 0, 1)/{\sqrt{2}}$.
As a result,
the effective Hamiltonian is obtained to be
\begin{align} 
\mathcal{H}_{\textrm{eff}}
	 \label{eq:Floquet-9} 
	 & = \mathcal{H}_0  + \delta m_0 \tau^z s^0
	 	- \mathcal{L}_y k_x \tau^x s^0,
\end{align}
with
\eq{
 \label{eq:Floquet-11} 
\mathcal{L}_{a=x,y,z} & =
\frac{ive^2}{m\hbar \Omega^3} (\bm{\mathcal{E}}\times \bm{\mathcal{E}}^*)_a.
}

It is noticed that the light-induced term $\mathcal{L}_y k_x \tau^x s^0$ in Eq. (\ref{eq:Floquet-9}) indicates the interaction depending on the spin angular momentum of light ($\sigma_{\textrm{L}}^z$),
orbital degrees of freedom ($\tau^x$), and the momentum $\bm{k}$.
The magnitude of the coefficient $\mathcal{L}_y$ is proportional to the laser intensity and $\Omega^{-3}$.
The sign of $\mathcal L_y$ denotes the chirality and the propagation direction of the light.

\subsection{ Light  along the $x$ axis}\label{sec:Floquet-x} 

The effective Hamiltonian for the light along the $x$ axis is derived by the $\pi/2$ rotation of that along the $y$ axis.
 $\bm{A}_{\textrm{L}}$ is represented by
\eq{
\bm{A}_{\textrm{L}}  = A_{\textrm{L}} (0, \cos \Omega t, \sigma_\textrm{L}^z \sin\Omega t).
}
The off-diagonal terms $\mathcal{H}_{0,-1}$ ($=\mathcal{H}_{1,0}^\dagger$) and $\frac{1}{\hbar \Omega}[\mathcal{H}_{0,-1}, \mathcal{H}_{0,1}]$ are obtained to be
\eq{ \nt
&\mathcal{H}_{0,-1}   =
		\frac{i eE_{\textrm{L}}}{2\hbar\Omega}  \left( \begin{matrix}
 		-\frac{\hbar^2}{m}
		[k^y + i\sigma_{\textrm{L}}^z k^z ]
		&
		-\sigma_{\textrm{L}}^z v
		\\
		\sigma_{\textrm{L}}^z v
		&
		\frac{\hbar^2}{m}
		[k^y + i\sigma_{\textrm{L}}^z k^z ]
		\end{matrix}\right),
\\
& \frac{1}{\hbar \Omega}[\mathcal{H}_{0,-1}, \mathcal{H}_{0,1}]
	 = 	- \mathcal{L}_x   \tau^x s^0 k^y.
}
Therefore, the effective Hamiltonian in the Floquet state is given by
\begin{align}   \nt
\mathcal{H}_{\textrm{eff}}  (\bm{k})
	 & = \mathcal{H}_0 + \delta m_0 \tau^z s^0
	 	- \mathcal{L}_x   k_y \tau^x s^0 .
\end{align}
It is noticed that the third term denotes the coupling between $\tau^x$ and $k_y$.

\begin{figure}[bt]\centering
\includegraphics[scale=0.45]{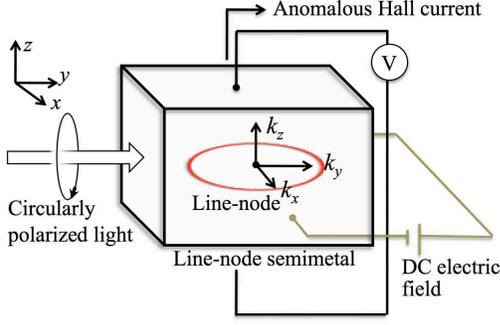}
\caption{(Color online)
Schematic illustration of a setup for the photovoltaic Hall effect in a line-node semimetal.
The incident light travels along the $y$ direction and the DC electric field is applied along the $x$ direction.
The photovoltaic anomalous Hall current flows along the $z $ direction.
}
\label{fig:fig1}
\end{figure}
\section{Photovoltaic anomalous Hall effect}\label{sec:PAHE} 
Using the effective Hamiltonian of Eq. (\ref{eq:Floquet-9}),
we will consider the photovoltaic anomalous Hall effect, which is a characteristic transport in the Floquet state, in the presence of both the incident circularly polarized light and an applied DC electric field.
The direction of the incident light and the applied electric field are along the $y$ and $x$ directions, respectively, as illustrated in Fig. \ref{fig:fig1}.
It is noted that before applying the circularly polarized light,
there is no anomalous Hall effect.

\begin{figure}[b]\centering
\includegraphics[scale=.42]{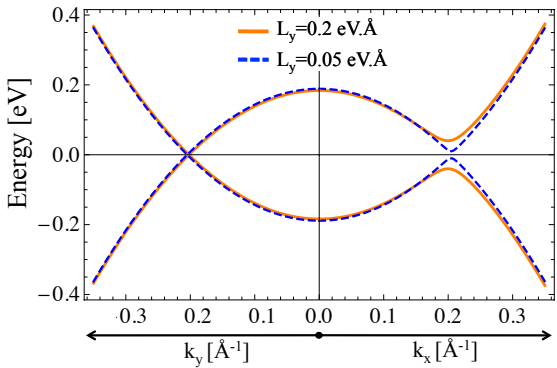}
\caption{(Color online)
The energy dispersion of the line-node semimetal in the Floquet state
at $k_z=0$ and $\epsilon_{\textrm{F}}=0$ for two finite $\mathcal{L}_y$.
The realistic parameters, $\hbar^2/(2m) = 4.5$ eV$\cdot$\AA$^2$, $v=2.5$ eV$\cdot$\AA, and $m_0= 0.184$ eV for Ca$_3$P$_2$, are used \cite{rf:Chan2016}.
(Left panel)
The position of the Weyl point is slightly shifted by the change of $m_0 \to \bar{m}_0= m_0 - \delta m_0$.
(Right panel)
A finite $\mathcal{L}_y$ plays a role to open the band gap,
and the line-node is vanished. 
\label{fig:dispersion}
}
\end{figure}

The energy spectrum of Eq. (\ref{eq:Floquet-9}) is obtained to be
\begin{align}
 E_\pm(\bm k)
 = -\epsilon_{\rm F} \pm \sqrt{
   \left( \frac{\hbar^2 k^2}{2m} - \bar m_0
   \right)^2
   + v^2 k_z^2
   + \mathcal L_y^2 k_x^2
 },
\end{align}
which hosts Weyl points at $k_x=k_z=0$ and $k_y=k_0$;
\begin{align} \label{eq:Weyl-point}
 k_0 = \frac{\sqrt{2 m \bar m_0}}{\hbar}.
\end{align}
The position of the Weyl point $k_0$ is slightly shifted by the change of $m_0 \to \bar{m}_0= m_0 - \delta m_0$
as shown in Fig. \ref{fig:dispersion}.
It is known that Weyl semimetals show anomalous Hall effect \cite{rf:Yang, rf:Burkov14}, then the present system under the light also shows it, as explained below.

\begin{figure}[bt] \centering
\includegraphics[scale=.132]{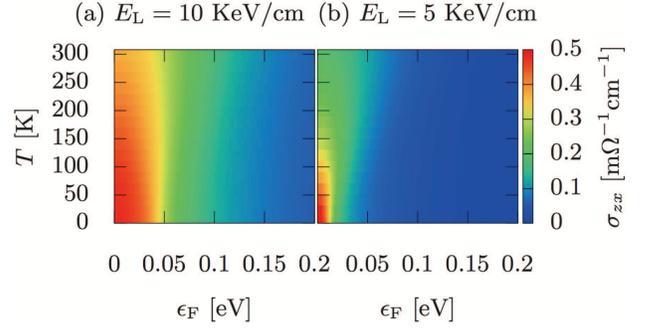}
\caption{(Color online)
Anomalous Hall conductivity $\sigma_{zx}^{\rm AHE}$ for (a) $E_{\rm L} = 10$ KeV/cm and for (b) $E_{\rm L} = 5$ KeV/cm, as functions of the Fermi level $\epsilon_{\rm F}$ and of the temperature $T$.
This figure takes the realistic parameters for Ca$_3$P$_2$ \cite{rf:Chan2016}, which are the same as Fig. \ref{fig:dispersion}. 
The light frequency is set to $\Omega = 100/(2\pi)$ THz.
The light-induced terms are evaluated as follows.
(a) $\delta m_0 = 0.82$ meV, $\mathcal L_y = 0.2$ eV$\cdot$\AA \ for $E_{\rm L} = 10$ KeV/cm and (b) $\delta m_0 = 0.21$ meV, $\mathcal L_y = 0.05$ eV$\cdot$\AA \ for $E_{\rm L} = 5$ KeV/cm, respectively.
These two cases (a) and (b) correspond to the solid and dashed lines in Fig. \ref{fig:dispersion}, respectively.
}
\label{fig:sigmaxy}
\end{figure}

\begin{figure}
\centering
\includegraphics[scale=.17]{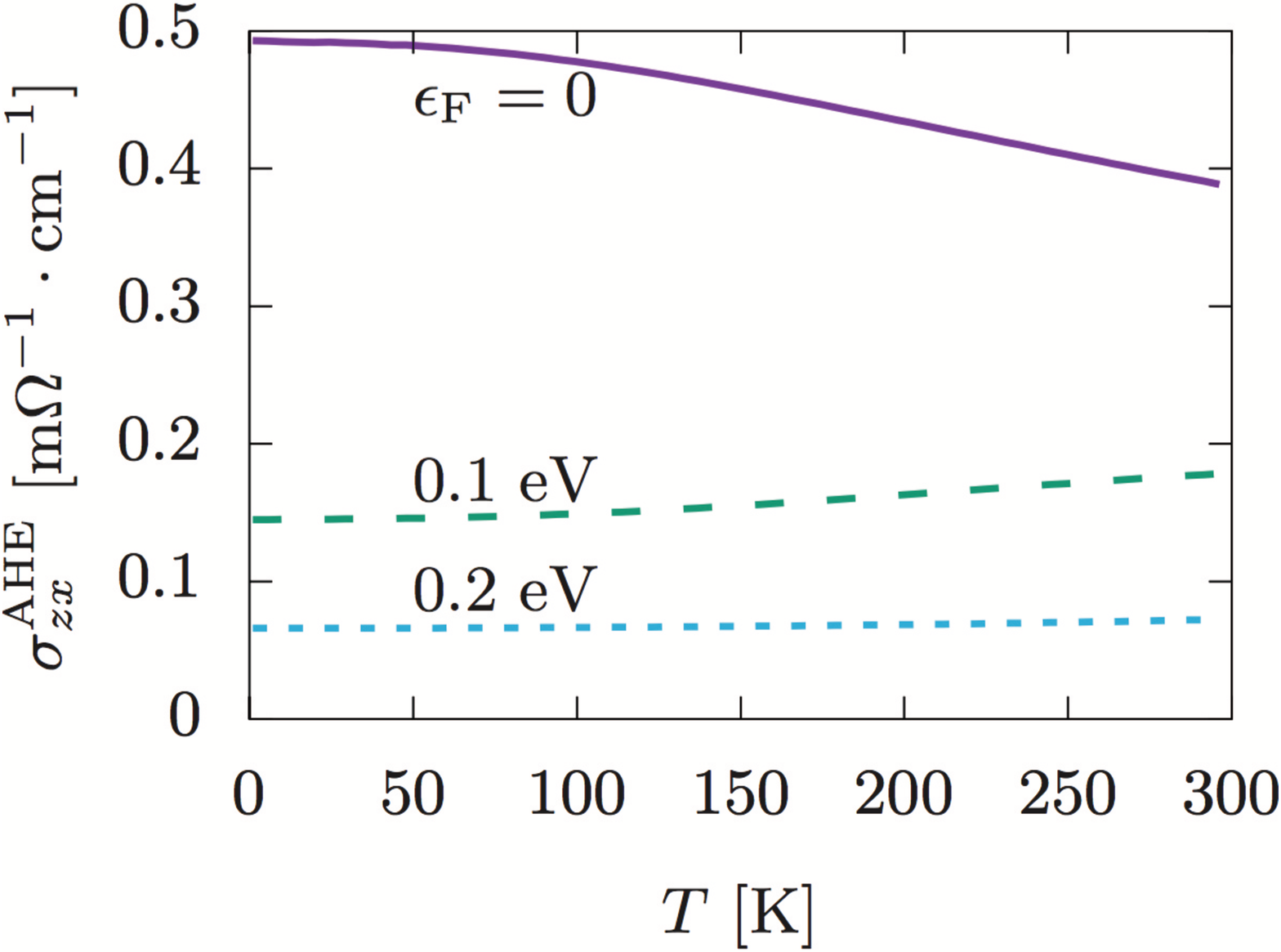}
\caption{(Color online) Temperature dependence of anomalous Hall conductivity for $\epsilon_{\rm F} = 0$ (solid line), $\epsilon_{\rm F}=0.1$ eV (dashed line), and $\epsilon_{\rm F} = 0.2$ eV (dotted line).}
\label{fig:sigmaxy-T}
\end{figure}

The Hall conductivity $\sigma_{zx}^{\rm AHE}$ is obtained by using the Kubo formula;
\begin{align}
&
 \sigma_{zx}^{\rm AHE}
 = - i \hbar e^2 \int \frac{d^3 k}{(2\pi)^3}
 \sum_{\alpha \ne \beta}
  \left[ v_{z}(\bm k) \right]_{\alpha \beta}
 \left[ v_x(\bm k) \right]_{\beta \alpha}
 \nonumber\\ & \times
 \frac{f \left(\epsilon_{\rm F} + E_\alpha(\bm k) \right) - f\left(\epsilon_{\rm F} - E_\beta(\bm k)\right)}
 {\left[E_\alpha(\bm k) - E_\beta(\bm k) \right]^2},
\end{align}
where $f$ is the Fermi distribution function and $v_i = (\partial H(\bm k) / \partial k_i)/\hbar$
is the velocity matrix along the $i$-th axis.
$\sigma_{zx}^{\rm AHE}$ is easily obtained for $\epsilon_{\rm F} = 0$ in the zero temperature $T = 0$.
Since
the system is a two-dimensional insulator for a fixed $k_y$, except for $k_y = k_0$, the two-dimensional Hall conductivity is quantized to $e^2/h \times \mathbb Z$.
The Hall conductivity in the whole system is given by the integral, i.e.,
\begin{align}
 \sigma_{zx}^{\rm AHE}|_{\epsilon_{\rm F} = T = 0}
 = 2 \times \frac{e^2}{h}
 \frac{k_0}{\pi},
 \label{eq:sigma0}
\end{align}
where the prefactor 2 comes from the spin degeneracy.
This is the same as that in a Weyl semimetal in which the Weyl points are located at $k_y=\pm k_0$ and $k_x=k_z=0$.
Note that the above value depends only on the location of the Weyl points, irrespective of the induced time-reversal-symmetry-breaking term $\mathcal L_y$.
For arbitrary $\epsilon_{\rm F}$ and $T$, $\sigma_{zx}^{\rm AHE}$ is numerically calculated, as shown in Fig. \ref{fig:sigmaxy}.
For $\epsilon_{\rm F} \to 0$ and $T \to 0$, the value of $\sigma_{zx}^{\rm AHE} = 2 e^2/h \times k_0/\pi = 0.5$ $\mathrm m\Omega^{-1} \cdot \textrm{cm}^{-1}$ is reproduced for both cases shown in Figs. \ref{fig:sigmaxy}(a), \ref{fig:sigmaxy}(b), and the solid line in Fig. \ref{fig:sigmaxy-T}.
The Hall conductivity  is reduced by the effects of finite temperature and Fermi energy, except for quite low temperature and Fermi-energy regime.

The maximum of $\sigma_{zx}^{\rm AHE}$ (Eq. \ref{eq:sigma0}) takes a large value for a large radius of the line node ($m$) and small $|E_{\rm L} / \Omega |$, because of Eqs. (\ref{eq:Floquet-12}) and (\ref{eq:Weyl-point}).
For finite $\epsilon_{\rm F}$ and $T$, on the other hand, $\mathcal L_y$ plays a important role: $\sigma_{zx}^{\rm AHE}$ is immediately suppressed when $\mathcal L_y$ is weak since $\mathcal L_y$ is proportional to the magnitude of the light-induced gap, which is essential for stability of the Hall conductivity for $\epsilon_{\rm F} \ne 0$ and $T \ne 0$.

Figure \ref{fig:sigmaxy-T} shows the temperature dependence of $\sigma_{zx}^{\rm AHE}$ for  three typical cases: (i) $\epsilon_{\rm F} = 0$; the Fermi level is located just on the Weyl points, (ii)  $\epsilon_{\rm F} = 0.1$ eV; carrier is moderately doped and the Fermi surface forms a torus, and (iii) $\epsilon_{\rm F} = 0.2$; carrier is sufficiently doped and a conventional sphere Fermi surface is realized.
The Hall conductivity $\sigma_{zx}^{\rm AHE}$ monotonically and algebraically decreases as a function of $T$ for $\epsilon_{\rm F} = 0$, while it slightly increases for finite Fermi energy ($\epsilon_{\rm F}= 0.1$ eV and 0.2 eV) in low-temperature regime and decreases to zero in the high-temperature limit (not shown in the figures).

\section{Discussion and Conclusion}
We will discuss an experimental realization of the circularly polarized light-induced effective Hamiltonian $\mathcal{H}_{\textrm{eff}}$ and nonzero $\sigma_{zx}^{\textrm{AHE}}$. 
The obtained result is verified when the energy scale of the incident light ($\hbar \Omega$) is much larger than that of the low-energy effective Hamiltonian $\mathcal{H}_0$,
i.e., $\hbar \Omega \gg \epsilon_{\textrm{F}}$.
The light along the in-plane ($xy$) direction breaks the mirror-reflection symmetry that protects the line-node on the $k_x k_y$ plane, hence the band gap opens except at the Weyl points.
As a result,
the nonzero anomalous Hall current, which is a characteristic transport in the Floquet states, is driven by
an applied DC electric field on the $xy$ plane, where the line node is located and perpendicular to the light direction.

It is found that from Fig. \ref{fig:sigmaxy},
the magnitude of $\sigma_{zx}^{\textrm{AHE}}$ is strongly reduced by a finite $\epsilon_{\textrm{F}}$,
and a giant $\mathcal{L}_y$ is needed for the detection of a nonzero $\sigma_{zx}^{\textrm{AHE}}$ in doped line-node semimetals.
Recently, a way to generate the giant electric field over $1$MV/cm in the infrared frequency (72 THz) \cite{rf:Sell2008} as well as terahertz (1THz) \cite{rf:Hirori2011} has been reported.
Such a light induces a giant orbital-momentum coupling $\mathcal{L}_y \approx$ 0.1 eV.\AA, which enables one to observe a large value of $\sigma_{zx}^{\textrm{AHE}}$ in a wide $\epsilon_{\rm F}$ and $T$ region [see Fig. \ref{fig:sigmaxy}(a)].
%

In a thin film of line-node semimetal, the Fermi level can be manipulated by using a gate voltage.
Thin film is suitable also for the irradiation of light, i.e.,
the light extends over the entire system when the thickness of the film is smaller than the wavelength of the light.
For example, one can use a 1$\mu$m (or thinner) film for a visible light, whose energy is large ($\hbar \Omega \sim 1$ eV $\gg \epsilon_{\rm F}$).
%
%
In addition, the anomalous Hall conductivity can be detected even if the intensity of the light is weak since
$\sigma_{zx}^{\textrm{AHE}}$ takes a giant value for $\epsilon_{\rm F} \sim 0$ as shown in Fig. \ref{fig:sigmaxy}.
It is also noticed that from Eq. (\ref{eq:sigma0}),
the anomalous Hall conductivity in the line-node semimetal is nearly independent of the laser intensity, and it is unlike that in Weyl semimetals: 
In the Weyl semimetal with the Floquet state,
the anomalous Hall conductivity depends on the laser intensity,
because the Weyl nodes are shifted by the circularly polarized light and its shift is proportional to the laser intensity \cite{rf:Oka15,rf:Chan15,rf:Taguchi16}.
In addition,
it is found that, unlike the light-induced effective Hamiltonian in the Dirac/Weyl semimetals,
this result is highly anisotropic for the direction of the incident light.

Finally, we consider applications of the photovoltaic anomalous Hall effect to optical and electrical devices.
In the presence of a weak light,
$\sigma_{zx}^{\textrm{AHE}}(\epsilon_{\textrm{F}})$ is about zero except for $\epsilon_{\textrm{F}}=0$,
so that there are giant difference between
$\sigma_{zx}^{\textrm{AHE}} (\epsilon_{\textrm{F}}\neq0)$
and
$\sigma_{zx}^{\textrm{AHE}} (\epsilon_{\textrm{F}}=0)$.
The Fermi level $\epsilon_{\rm F}$ is controlled by the on-off gate voltage in a thin film.
From the analogy to the resistivity random access memories or the phase random access memories,
the magnitude of the Hall conductivity $\sigma_{zx}^{\textrm{AHE}} (\epsilon_{\textrm{F}}\neq0)$
and $\sigma_{zx}^{\textrm{AHE}} (\epsilon_{\textrm{F}}=0)$ can be regarded as the signal "0" and "1".
Such a giant difference of the Hall conductivity
can be applicable to optical and electrical memory devices or phototransistor.
Its basic mechanism could lie on the characteristic property of line-node semimetals.
Additionally, it is noticed that our
 anomalous Hall effect can be useful for the detection of the direction of light,
because the nonzero $\sigma_{zx}^{\textrm{AHE}}$ appears when the light is along the mirror-symmetry-breaking direction of the line-node semimetal.


\textit{Note added}.--- During preparation of the manuscript, we became aware of similar works by
Z. Yan \textit{et. al.,} \cite{rf:Z-Yan2016}, 
C.-K. Chan \textit{et. al.,}\cite{rf:C-Chan2016}, 
and
A. Narayan \cite{rf:Narayan2016}, 
which discuss the transition from a line-node semimetal into Weyl semimetals.



\section*{Acknowledgments}
The authors acknowledge the fruitful discussion with Y. Tanaka.
This work was supported by Grants-in-Aid for the Core Research for Evolutional Science and Technology (CREST) of the Japan Science, for Japan Society for the Promotion of Science (JSPS) Fellows,
for Challenging Exploratory Research (Grant No.15K13498), and by Grant-in-Aid for Young Scientists B (Grant No. 16K17725).

\appendix

\section{Symmetry consideration for effective Hamiltonian induced by circularly polarized light}
\label{symmetry}

The form of effective Hamiltonian in the Floquet state is determined by the symmetry consideration, as shown below.

\subsection{Symmetry of circularly polarized light}

Circularly polarized light along $\bm q$ is obtained from the vector potential of
\begin{align}
  \bm A_{\rm L} = R(\hat{\bm q}) A_{\rm L} (\cos \Omega t, \sigma_{\rm L}^z \sin \Omega t, 0)^{\rm T},
  \
  A_{\rm L} \in \mathrm i \mathbb R,
\end{align}
where  $R(\bm q) \in \rm SO(3)$ is the rotational matrix that rotates $\hat{\bm z}$ into $\bm q$.
The Hamiltonian describing the interaction between an electronic system and the light has the form
\begin{align}
 H_{\rm em}(t) = - \bm J \cdot \bm A_{\rm L},
\end{align}
where $\bm J$ denotes the charge current of the system.

\begin{figure}
\centering
\includegraphics[scale=0.5]{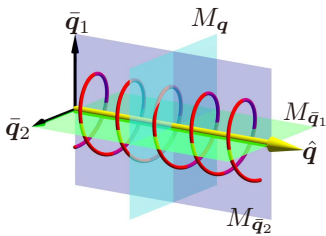}
\caption{Circularly polarized light along $\hat{\bm q}$ and mirrors.}
\label{light}
\end{figure}

$H_{\rm em}(t)$ has mirror-reflection $M_{\bm q}$ symmetry perpendicular to $\bm q$  and rotational symmetry along ${\bm q}$.
As a result, spatial-inversion symmetry is also respected.
$\bm J$ is time-reversal odd.
Time-reversal $\mathcal T$, on the other hand, reverses the chirality of the light, i.e., $H_{\rm em}(t)$ breaks time-reversal symmetry.
The two mirrors $M_{\bar{\bm q}_1}$ and $M_{\bar{\bm q}_2}$ parallel to ${\bm q}$, which is shown in Fig. \ref{light}, also reverses the chirality.
Therefore, $H_{\rm em}(t)$ is invariant for the composite operation, so-called magnetic reflection, $M_{\bar{\bm q}_i} \mathcal T$;
\begin{align}
 (M_{\bar{\bm q}_i} \mathcal T)^{-1} H_{\rm em}(t) (M_{\bar{\bm q}_i} \mathcal T) = H_{\rm em}(-t).
\end{align}
Similarly, one can see that magnetic-rotational $C(\pi {\bar{\bm q}}_i) \mathcal T$ symmetry along ${\bm q}$ holds.

In summary, $H_{\rm em}(t)$ is invariant against the following symmetry operations; $C(\theta \hat{\bm q})$, $M_{\bm q}$, $I$, $M_{\bar{\bm q}_i} \mathcal T$, $C(\pi {\bar{\bm q}}_i) \mathcal T$.
For the realization of the photovoltaic Hall effect, the Floquet effective Hamiltonian must have a time-reversal-odd term of the $A_{2g}$ representation or its compatible ones.

\subsection{Minimal model for a line-node semimetal}

A minimal ($D_{\infty h}$) model hosting a line node consists of even-- ($\tau^z=+1$) and odd-- ($\tau^z = -1$) parity orbitals under the mirror reflection with respect to the horizontal plane.
The symmetry operations, spatial inversion $I$, mirror reflection with respect to the $x_i=0$ plane $M_{i}$, are represented by
\begin{align}
 I = \tau^z,
 \
 M_x = s^x,
 \
 M_y = s^y,
 \
 M_z = \tau^z s^z.
\end{align}
The 16 matrices $\tau^\mu \sigma^\nu$ in this theory are summarized in Table \ref{matrices}.
\begin{table}
\centering
\caption{16 matrices in the minimal theory.}
\begin{tabular}{cccccc}
 \hline\hline
  & $I$ & $M_x$ & $M_y$ & $M_z$ & $\mathcal T$
 \\
 \hline
  $\tau^0 s^0$, $\tau^z s^0$ & $+$ & $+$ & $+$ & $+$ & $+$
  \\
 $\tau^0 s^x$, $\tau^z s^x$ & $+$ & $+$ & $-$ & $-$ & $-$
 \\
 $\tau^0 s^y$, $\tau^z s^y$ & $+$ & $-$ & $+$ & $-$ & $-$
 \\
 $\tau^0 s^z$, $\tau^z s^z$ & $+$ & $-$ & $-$ & $+$ & $-$
 \\
 $\tau^x s^0$ & $-$ & $+$ & $+$ & $-$ & $+$
 \\
 $\tau^x s^x$ & $-$ & $+$ & $-$ & $+$ & $-$
 \\
 $\tau^x s^y$ & $-$ & $-$ & $+$ & $-$ & $-$
 \\
 $\tau^x s^z$ & $-$ & $-$ & $-$ & $-$ & $-$
 \\
 $\tau^y s^0$ & $-$ & $+$ & $+$ & $-$ & $-$
 \\
 $\tau^y s^x$ & $-$ & $+$ & $-$ & $+$ & $+$
 \\
 $\tau^y s^y$ & $-$ & $-$ & $+$ & $+$ & $+$
 \\
 $\tau^y s^z$ & $-$ & $-$ & $-$ & $-$ & $+$
 \\ \hline
 $k_x$ & $-$ & $-$ & $+$ & $+$ & $-$
 \\
  $k_y$ & $-$ & $+$ & $-$ & $+$ & $-$
  \\
   $k_z$ & $-$ & $+$ & $+$ & $-$ & $-$
 \\
 \hline\hline
\end{tabular}
\label{matrices}
\end{table}

A Hamiltonian for a line-node semimetal, which is $A_{1g}$ representation, is given by
\begin{align}
 H_0(\bm k) = c(\bm k) \tau^0 s^0
 + m(\bm k) \tau^z s^0
 + v k_z \tau^y s^0,
\end{align}
with
\begin{align}
 c(\bm k) &= c_0 + c_1 k_z^2 + c_2 (k_x^2 + k_y^2),
 \\
 m(\bm k) &= m_0 + m_1 k_z^2 + m_2 (k_x^2 + k_y^2).
\end{align}
In Eq. (\ref{eq:2-1}), $c(\bm k)$ term is ignored and $m_1=m_2$ is assumed, for simplicity.

In the following, we derive the effective Hamiltonian from the symmetry consideration.
The Floquet Hamiltonian shares the same symmetry as $H_0(\bm k) + H_{\rm em}(t)$.
$H_{\rm em}(t)$ renormalizes the parameters in $H_0(\bm k)$ and yields new terms $\tilde H_{\rm em}(\bm k)$.
Here we focus time-reversal-symmetry-breaking terms in $\tilde H_{\rm em}(\bm k)$, which may trigger the photovoltaic anomalous Hall effect.

\subsubsection{Circularly polarized light along the $z$ direction}

Symmetry of the system under the circularly polarized light along the $z$ direction is down to $\infty/mm'm'$ from $\infty/mmm$, where $'$ denotes the magnetic operation.
The time-reversal-symmetry-breaking terms induced by the light belong to the time-reversal-odd $A_{2g}$ representation of $\infty/mmm$;
\begin{align}
 \tilde H_{\rm em} &= (a_{0z} \tau^0 + a_{zz} \tau^z) s^z
 + \lambda_{y} \tau^y (k_x s^x + k_y s^y)
 \nonumber\\ & \quad
  + a_{yz} k_z \tau^y s^z + \mathcal O(k^2).
\end{align}
The light may induce many spin-dependent terms only when the system has spin-orbit interaction.

\subsubsection{Along the $x$ direction}
Symmetry of the system under the circularly polarized light along the $x$ direction becomes $m'mm'$ from $\infty/mmm$.
The time-reversal-symmetry-breaking terms are the time-reversal-odd $E_{(2n-1)g}$ representation of $\infty/mmm$;
\begin{align}
 \tilde H_{\rm em} &= (a_{0x} \tau^0 + a_{zx} \tau^z ) s^x + \tau^y (a_{yz} k_x s^z + a_{yx} k_z s^x )
 \nonumber\\ & \quad
 +
 a_{x0} k_y \tau^x s^0
 + \mathcal O(k^2).
\end{align}
Similarly to the previous case under the light along the $z$ direction, many spin-dependent terms may be induced in the presence of spin-orbit interaction.
Note that
the last term shows up even in the absence of spin-orbit interaction since it is independent of spin.

\end{document}